# A Horizontal Vane Radiometer: Experiment, Theory, and Simulation

David Wolfe[1], Andres Larraza[1,a], and Alejandro Garcia[2]

[1]*Department of Physics, Naval Postgraduate School, Monterey CA 93940, USA*

[2]*Department of Physics and Astronomy, San Jose State University, San Jose CA 95152, USA*

The existence of two motive forces on a Crookes radiometer has complicated the investigation of either force independently. The thermal creep shear force in particular has been subject to differing interpretations of the direction in which it acts and its order of magnitude. In this article we provide a horizontal vane radiometer design which isolates the thermal creep shear force. The horizontal vane radiometer is explored through experiment, kinetic theory, and the Direct Simulation Monte Carlo (DSMC) method. The qualitative agreement between the three methods of investigation is good except for a dependence of the force on the width of the vane even when the temperature gradient is narrower than the vane which is present in the DSMC method results but not in the theory. The experimental results qualitatively resemble the theory in this regard. The quantitative agreement between the three methods of investigation is better than an order of magnitude in the cases examined. The theory is closer to the experimental values for narrow vanes and the simulations are closer to the experimental values for the wide vanes. We find that the thermal creep force acts from the hot side to the cold side of the vane. We also find the peak in the radiometer's angular speed as a function of pressure is explained as much by the behavior of the drag force as by the behavior of the thermal creep force.

---

[a] Author to whom correspondence should be addressed. Electronic mail: larraza@nps.edu

# I. INTRODUCTION

Thermal creep is the flow of a gas over a surface due to a temperature gradient in the gas parallel to the surface. The force on the gas is equal and opposite to the force on the surface. The force exerted on a surface by a gas with a parallel temperature gradient is thus called the thermal creep force. The Einstein effect is also caused by a temperature gradient in a gas, but it acts on surfaces perpendicular to the temperature gradient. It occurs when the structure is thin in the direction of the temperature gradient and is only experienced on those areas of the surface within a mean free path of the edge. These forces are most famously demonstrated by the Crookes radiometer. A thorough review of radiometric phenomena including historical accounts can be found in reference [1]. Figure 1 shows where the two forces act on the vanes of a Crookes radiometer.

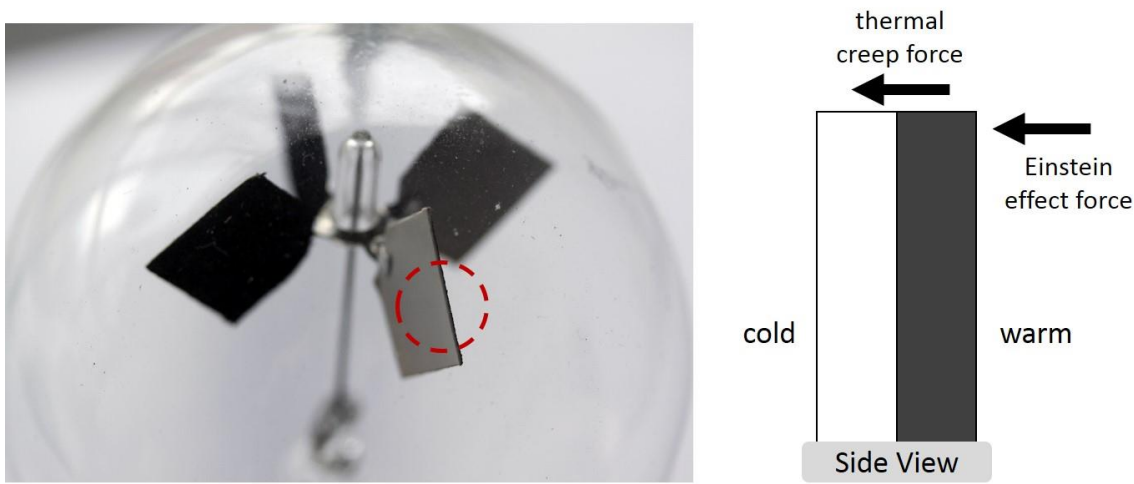

Fig 1. Motive forces on the vane of a Crookes radiometer. The figure on the right is an edge view of the vane and not to scale. In commercial Crookes radiometers, typical vanes are 1.5 cm x 1.5 cm and thickness is usually on the order of 0.1mm. The thermal creep force acts on the edge parallel to the temperature gradient. The Einstein effect acts on the surfaces perpendicular to the temperature gradient.

The Hettner radiometer was designed to measure the shear pressure or force on a plate with a uniform temperature from the thermal creep caused by a parallel plate with a temperature gradient. [2] This design does not experience the Einstein effect because the plate face is parallel to the temperature gradient. Figure 2 shows the forces on a Hettner radiometer.

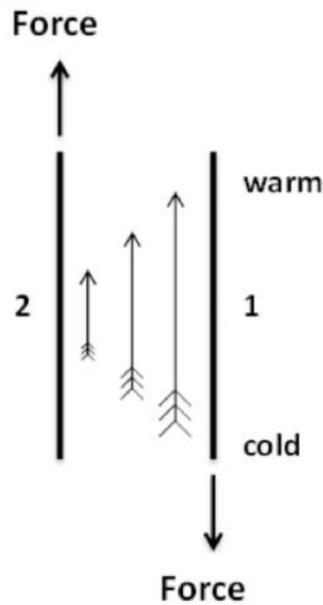

Fig. 2. Hettner Radiometer. Plate 1 has a temperature gradient and plate 2 has a uniform temperature. The temperature gradient establishes thermal creep (arrows with fletchings).

We designed a horizontal vane radiometer shown in Figure 3, which like the Hettner radiometer does not experience the Einstein effect because the temperature gradient is parallel to the vane face but which like the Crookes radiometer is free to rotate around a spindle.

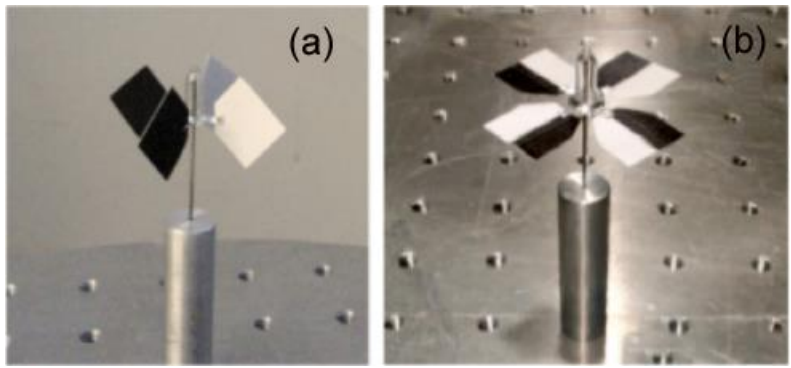

Fig. 3. The Crookes and Horizontal Vane Radiometers. (a) The Crookes radiometer experiences thermal creep forces and the Einstein effect. (b) The horizontal vane radiometer experiences thermal creep forces but no Einstein effect.

## II.    EXPERIMENT

The angular speeds of horizontal vane radiometers were observed over a range of pressures with mean free paths from 5 mm to 0.2 mm. We report results on two types of horizontal vane radiometers. The narrow vane radiometer had 8 mm by 16 mm vanes. The wide vane radiometer had 16 mm by 16 mm vanes. The vanes were constructed from high gloss photo paper. Half of the non-glossy side was printed with black LaserJet ink. Two pieces of paper cut into the shape of the vanes were glued together glossy side to glossy side so that both the upper and lower horizontal surfaces would have a white side and a black side. The vanes on both radiometers were connected to the spindle with 4 mm stems.

A high intensity radiant heat projector was used to illuminate the radiometers. The temperature profile that developed on the vanes when illuminated was observed with a FLIR SC8200 camera outside of the vacuum chamber and a FLIR ThermoVision A20 camera inside of the vacuum chamber. The temperature profile was not observed to be sensitive to pressure. Under illumination, the vanes developed a 9 K temperature difference between the black and the white side with the transition in temperature occurring mostly in a sharp gradient in the center of the vane with a characteristic length of 3.5 mm. The radiometers were tested in a 18 inch diameter 12 inch high Pyrex bell jar on a highly reflective test stand so that both the upper and lower horizontal surfaces would be illuminated. Figure 4 reports the results of these experiments at the pressures tested. [3] Reference [4] shows video of horizontal vane radiometers rotating at 1.33 Pa. The video shows various radiometer configurations and dimensions. Please note the dimensions of the radiometers in the video are generally different than the dimensions of the radiometers reported in this experiment.

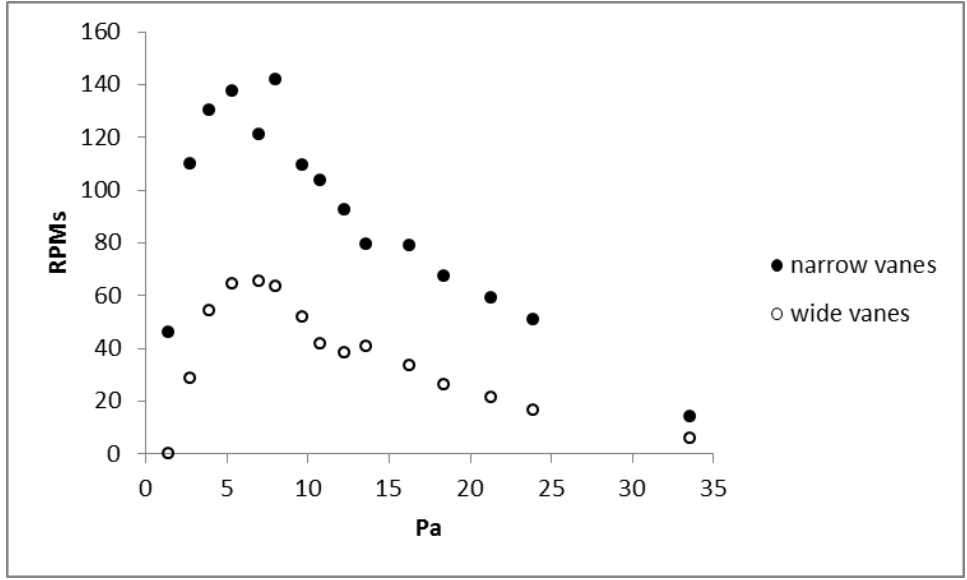

Fig 4. Angular Speed (experiment). The narrow vanes are 8mm by 16mm and the wide vanes are 16mm by 16mm. Both radiometers have four vanes. Vanes of both radiometers have a 4mm stem connecting them to the spindle axis. The temperature difference on the vanes is 9K. The characteristic length of the temperature gradient is 3.5mm. The observed motion is with the white (cold) side leading. The size of the data points is about the extent of experimental error.

## III.    THEORY

Thermal creep explains why the horizontal vane radiometer rotates, but an understanding of why the angular speed peaks and is a function of the vanes' widths requires careful examination of the details of both the thermal creep shear force and the drag force on the vanes.

### A.    THERMAL CREEP

For an angular speed of 10 radians per second the furthest edges of the vanes will move a distance between 2 μm and 80 nm during a mean free time between collisions for the pressures in our experiment. Because these distances are small relative to the widths of the vanes in our experiment, a temperature gradient in the gas over the vane created by collisions between the gas and the vane will remain over the vane despite the vane's rotation justifying a quasi-static assumption for the calculation of the thermal creep force on a moving radiometer with the observed velocities.

Scandurra *et al.* derive an analytical formula for the thermal creep shear stress using the Chapman-Enskog method, [5]

$$\Delta p_{TC} = \frac{15}{64\sqrt{2}} \frac{k_B}{\sigma_{CS}} \frac{dT}{dy} \alpha, \tag{1}$$

where $k_B$ is Boltzmann's constant, $\sigma_{CS} = \pi d^2$ is the hard-sphere collision cross section of the molecules with diameter d, dT/dy is the temperature gradient in the gas, and α is the accommodation coefficient. In Maxwell's kinetic theory of gas particle-surface interactions, the accommodation coefficient is the extent to which energy and momentum are transferred in a particle-surface collision. If all collisions are assumed to be either perfectly specular or perfectly diffuse, it is the fraction of diffuse collisions in a sample. The value of the accommodation coefficient thus ranges between zero and one. The Maxwell model of accommodation is adopted in Scandurra *et al.* and in the present work.

The thermal creep force $F_{TC}$ is this pressure multiplied by the area over which it acts $A_{TC}$. If $L_{TC}$ is the length of a rectangle perpendicular to the temperature gradient and $W_{TC}$ is the width along the temperature gradient, we have

$$F_{TC} = \Delta p_{TC} A_{TC} = \Delta p_{TC} L_{TC} W_{TC}. \tag{2}$$

Assuming the temperature gradient in the gas is roughly constant in the area where the thermal creep force acts, the temperature gradient term can be replaced by

$$\frac{dT}{dy} \approx \frac{\Delta T}{W_{Ggas}}, \tag{3}$$

where $\Delta T$ is the temperature difference between the hot and the cold sides of the surface and $W_{Ggas}$ is the width of the temperature gradient in the gas.

The width of the temperature gradient established in a low-pressure gas by a surface with a temperature gradient is wider than the temperature gradient on the surface $W_{Gsurf}$ by a length equal to the slip length, [6] or

$$W_{Ggas} = W_{Gsurf} + 2\lambda\frac{2-\alpha}{\alpha} \qquad (4)$$

where $\lambda$ is the mean free path of the gas. The mean free path is related to the particle number density n and collision cross section as $\lambda = 1/\sqrt{2}n\sigma_{CS}$ .

The physical meaning of equation (4) can be understood by considering the two extreme values of $\alpha$. When the accommodation equals zero, the width of the gradient in the gas is infinite. There is thus no gradient when there is no energy transfer between the surface and the gas. When the accommodation equals one, the width of the gradient in the gas is the width of the gradient on the surface plus two mean free paths. The gradient thus extends one mean free path further on each side in the gas than in the surface.

In the horizontal vane radiometer experiment, there are four temperature gradients in the plane of each vane shown in figure 5.

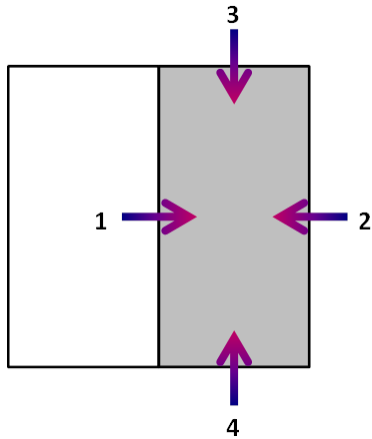

Fig 5. Temperature Gradients on a Horizontal Vane. The white (cold) side of the vane is assumed to have a temperature similar to ambient temperature and the grey (hot) side is warmer than ambient temperature.

Temperature gradient #1 causes the force $F_{TC1}$ which causes the rotation of the radiometer and is a convenient starting point for the calculation of the total thermal creep force on the vane. We will treat the effects of the other temperature gradients as apparent modifications of the force from temperature gradient #1.

Temperature gradient #2 creates a force which opposes the rotation. The magnitude of this force is less than the magnitude of the force created by temperature gradient #1 because the portion of temperature gradient #2 which is not over the vane does not create a force, i.e.

$$F_{TC} = xF_{TC1},$$ (5)

where x is a numerical factor between zero and one representing the fraction of the force from temperature gradient #1 which is offset by the force from temperature gradient #2.

The net force created by temperature gradients #3 and #4 is zero, but these gradients decrease the magnitude of temperature gradient #1 near the edges. This has the effect of decreasing the apparent length over which the net thermal creep pressure from temperature gradient #1 acts. The length correction is the slip length times the numerical factor β or

$$L_{TC}{}' = L_{TC1} - \beta\lambda\frac{2-\alpha}{\alpha}.$$ (6)

The width over which temperature gradient #1 acts is the smaller of either the width of the vane $W_V$ or the width of the temperature gradient in the gas. Putting this together with equations (1) to (6) gives the following expression for the net thermal creep force on the two sides of a single vane in the horizontal vane radiometer experiment:

$$F_{TC} = \frac{x15}{32\sqrt{2}} \frac{k_B}{\sigma_{CS}} \Delta T \alpha L' \min\left(\frac{W_V}{W_{Ggas}}, 1\right),$$  (7)

where $L'$ is the length of the vane with the slip length correction.

Figure 6 shows the net thermal creep force predicted by this equation for the vanes in the horizontal vane radiometer experiment assuming numerical factors of x = 0.5 and β = 0.5 for equations (5) and (6) and an accommodation coefficient of α = 0.6. The values of x and β follow from the assumption that the gas temperature at the edge is equal to the average of the temperature of the hot side of the vane and the ambient temperature. The accommodation value is the maximum value of the modified Boule formula for estimating clean surface accommodation coefficients. [7] We used the modified Boule formula because of the inability to find a tabulated value for the accommodation coefficient of paper in air. We chose the maximum value because the vanes were not "clean" and because the tabulated values that do exist for other materials often exceed the modified Boule formula estimate. An even higher value of the accommodation would result in the force rising more quickly as a function of pressure and asymptotically approaching a higher value. The pressure dependence of the force is a result of the slip length corrections in the width of the gradient in the gas and the apparent length of the vane.

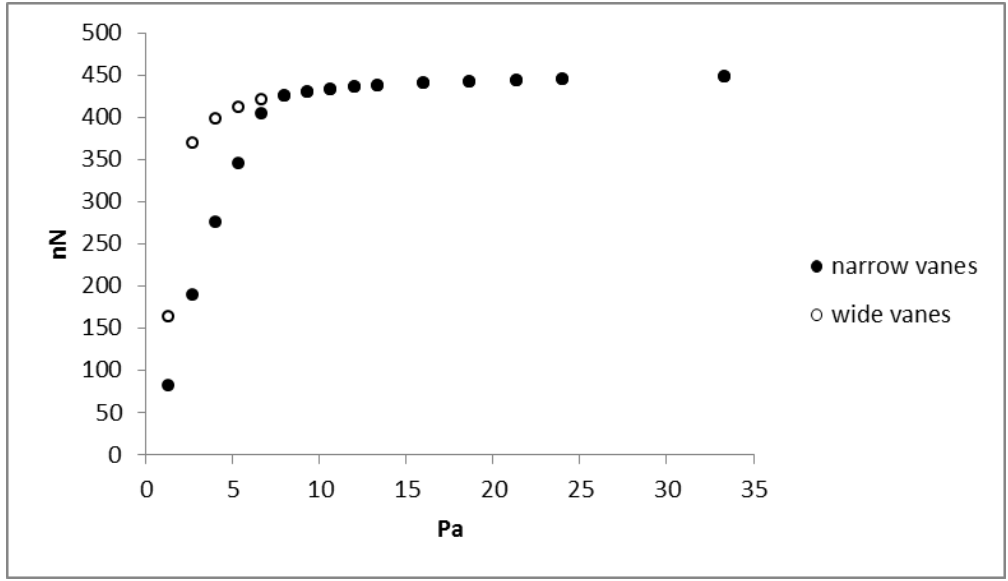

Fig 6. Thermal Creep Force (theory). The narrow vanes are 8mm by 16mm and the wide vanes are 16mm by 16mm. The temperature difference on both vanes is 9K. The characteristic length of the temperature gradient on the surface is 3.5mm. The accommodation coefficient is 0.6. At 8 Pa and greater pressure the forces are similar on the narrow and wide vanes. A positive force is directed from the hot side to the cold side.

## B.    DRAG FORCE

The dominant drag force on a thin plate oriented parallel to the flow is skin friction. The Reynolds numbers for every trial in the horizontal vane radiometer experiment was less than 0.02. Therefore, for all the trials in this experiment the flow can be regarded as being in the Stokes' regime. The drag coefficient for Stokes' drag is proportional to the inverse of the Reynolds number.

$$C_D = \frac{\delta \mu}{\rho U W},$$  (8)

where $\delta$ is a numerical factor, $\mu$ is the viscosity of the fluid, U is the characteristic flow velocity, W is the length over which the flow velocity varies (the width of the vane in this case), and $\rho$ is the density of the fluid.

For thin plates oriented parallel to the flow, a theoretical value of 4.12 for $\delta$ has been calculated. [8] On the other hand, an experimental value of 5.91 has been observed.

[9] Thus for uniform flow past a flat plate the Stokes drag force on both sides of the plate is

$$F_D = 2\frac{1}{2}\rho U^2 C_D LW = 5.91 U\mu L,$$
(9)

where L is the dimension perpendicular to the flow (the length of the vane in this case).

The Stokes' drag equation does not take edge effects into account. Each edge parallel to the flow that is immersed in the fluid adds the empirically determined correction to the drag force equation [10]

$$F_{edge} = 1.6 U\mu W.$$
(10)

At low pressure there is an apparent decrease in the dynamic viscosity due to the slip length [6]

$$\mu' = \frac{\mu W}{W + 2\lambda\frac{2-\alpha}{\alpha}}.$$
(11)

Chapman's expression for the dynamic viscosity, [11]

$$\mu = \frac{5\pi}{32}\rho C\lambda,$$
(12)

which accounts for variable collision rate, improves on Maxwell's classical estimate, $\mu = \rho C \lambda / 3$. Here C is the mean speed of a gas molecule. Note that μ is pressure independent. The substitutions ρ=nm and $C = \sqrt{8 k_B T / \pi m}$, where m is the molecular mass, express $\mu'$ in terms of kinetic theory quantities

$$\mu' = \frac{5\sqrt{\pi}}{16} \frac{\sqrt{m k_B T} W}{\sigma_{CS} W + \frac{\sqrt{2}}{n} \frac{2 - \alpha}{\alpha}}. \tag{13}$$

It is not obvious that the slip length correction should apply to the drag on the edges, but a correction is necessary for the edge drag force to equal zero at zero pressure. We will assume the correction takes a similar form for both the skin friction and the edge drag.

For a uniformly rotating vane, including edge effects, the drag force is

$$F_D = \omega \left( L_S + \frac{L}{2} \right) \mu' \left( 5.91 L + 3.2 W \right), \tag{14}$$

where $L_s$ is the length of the stem attaching the vane to the spindle. Figure 7 shows the drag force predicted by this equation for the vanes in the horizontal vane radiometer experiment with a 4 mm stem at 10 rad/s.

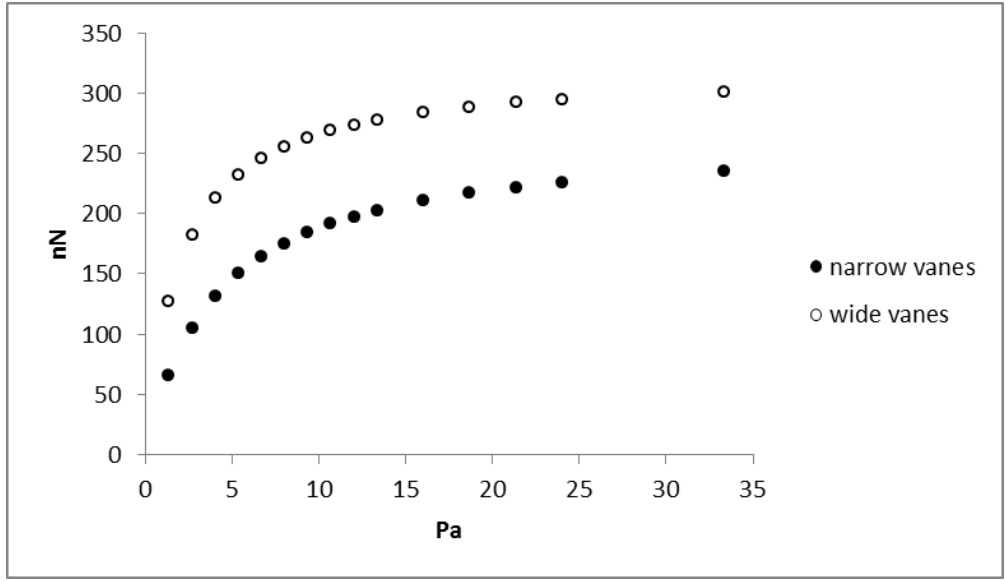

Fig 7. Drag Force at 10 rad/s (theory). The narrow vanes are 8mm by 16mm and the wide vanes are 16mm by 16mm. Both vanes have a 4mm stem connecting them to the spindle axis. The accommodation coefficient is 0.6.

## C.    ANGULAR SPEED

Along with our quasi-static assumption, we also made an estimate of the centrifugal effects by calculating the centrifugal force on gas particles within a volume comprised by a mean free path above and below the vane's surface and determined them to be less than a tenth of a percent of those due to thermal creep and drag forces.  Thus, assuming negligible mechanical friction in the spindle, the equilibrium angular speed $\omega$ of the horizontal vane radiometer occurs when the drag torque equals the thermal creep torque. Because the drag force is linearly related to the angular speed, we can define the drag force per unit angular velocity $f_D$ as

$$f_D = \frac{F_D}{\omega}.$$
(15)

Thus, in terms of thermal creep,

$$\omega = \frac{r_{tc} F_{TC}}{r_d f_D} \qquad (16)$$

where $r_{tc}$ and $r_d$ are the effective moment arms of the thermal creep and drag forces respectively. In our experimental geometry, the drag force moment arm is 121% and 125% of the thermal creep force moment arm for the narrow and wide vane respectively because the thermal creep force acts uniformly along the length of the vane while the drag force increases further from the rotation axis. Since both torques will scale linearly with the number of vanes it is possible to examine the torques on a single vane to predict the angular speed of a radiometer with four vanes. Figure 8 is the predicted angular velocity that results by using the expressions (14) and (7) in equation (16).

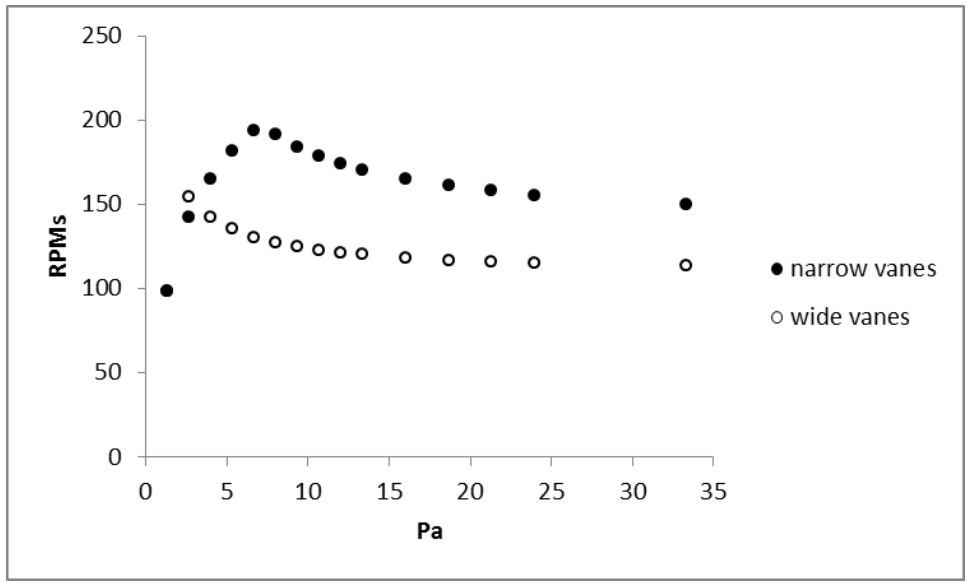

Fig 8. Angular Speed (theory). The narrow vanes are 8mm by 16mm and the wide vanes are 16mm by 16mm. Both vanes have a 4mm stem connecting them to the spindle axis. The temperature difference on both vanes is 9K. The characteristic length of the temperature gradient on the surface is 3.5mm. The accommodation coefficient is 0.6. Positive angular speed represents rotation with the cold side leading. The values at 1.33 Pa are nearly equal.

# IV. SIMULATIONS

Direct Simulation Monte Carlo (DSMC) is a method of determining continuum properties of a fluid by simulating individual particles of the fluid. Simulation particles are weighted to represent multiple fluid particles to reduce computational effort. SPARTA (Stochastic PArallel Rarefied-gas Time-accurate Analyzer) is an open-source DSMC program developed at Sandia National Laboratories. [12] [13] We used it in this research to determine separately the shear pressures caused by drag and thermal creep on vanes similar to those in the horizontal vane radiometer experiment at the range of pressures tested in the experiment. We used the February 21, 2015 version of SPARTA with one modification. The modification was the removal of the command to terminate the trial if a particle collides with an interior surface of the vane. This feature was in the code to ensure objects in the simulation space are airtight, but occasionally the algorithm used by SPARTA to distinguish between interior and exterior collisions mistakenly classified a collision in our geometry. We checked the simulated vanes for airtightness before running the trials on the modified program. We ran the trials on the Hamming supercomputer at the Naval Postgraduate School and the Spirit supercomputer at the Air Force Research Laboratory. Each trial took five to fifty-thousand CPU hours to complete.

## A.     THERMAL CREEP SIMULATIONS

We ran a total of 30 thermal creep simulations encompassing both the narrow vane and the wide vane at the 15 pressures tested in the experiment. In each trial, we simulated a single stationary vane and the shear pressure on the vane was the output. We divided each trial into 10 time segments which reported 10 separate average shear pressures. This confirmed the behavior was not transitory and allowed us to calculate a standard deviation between the 10 segments to quantify the uncertainty in the output.

In SPARTA, geometry objects have a single assigned temperature and cannot touch other objects. To simulate a vane with a temperature gradient, we created nine objects in the simulation space. Figure 9 is an overhead view of the wide and narrow vane as modeled in SPARTA.

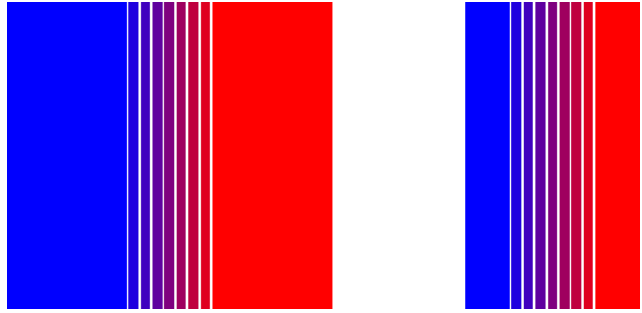

Fig. 9. Vanes as Modeled in SPARTA. The wide and narrow vanes are each divided into nine segments separated by two Angstroms. Each segment has a single assigned temperature which differs from the other segments.

The cold side of the vanes was assigned a temperature of 293 K. The hot side of the vanes was assigned a temperature of 302 K. Each intermediate segment was 1.125 K warmer than its colder neighbor. All nine rectangular boxes which comprised a vane were 16 mm long and 0.1 mm thick. The seven boxes in the gradient region were 0.5 mm wide which from equation (4) is thin enough to create a smooth gradient in the gas at all pressures simulated. The two end boxes were 6.25 mm wide for the wide vane and 2.25 mm wide for the narrow vane.

The simulation space was 18 mm by 26 mm by 10 mm for the narrow vane simulations and 26 mm by 26 mm by 10 mm for the wide vane simulations. This ensured at least one mean free path between the vane and the boundary at all pressures simulated. The boundary condition was inflow and outflow so particles whose trajectory took them outside the simulation space would no longer be simulated and new particles would be created and entered into the simulation space throughout the trial. The size of the grid cells was varied with pressure so that each edge of a grid cell equaled half of a mean free path in length. The total number of grid cells ranged from 280 for the narrow vane at 1.33 Pa to over 6 million for the wide vane at 33.33 Pa.

Particles were created with an initial average temperature of 293 K. 850 million particles were initially created in the narrow vane simulations and 1.2 billion particles were initially created in the wide vane simulations. This ensured there was an average of 200 particles per grid cell at the highest pressure trials. The lower pressure trails had more than 200 particles per grid cell since the grid cells were larger and fewer in the lower pressure trials. The number of molecules represented by a single simulation particle varied with pressure. This value varied from 1.8 million molecules per simulation particle at 1.33 Pa to 45 million molecules per simulation particle at 33.33 Pa. 79% of the particles were diatomic Nitrogen and 21% of the particles were diatomic Oxygen. We used the variable soft sphere collision model.

The time step was 100 ns. This value is one fourth the mean free time between collisions at 33.33 Pa and a smaller fraction of the mean free time between collisions at lower pressures. Each simulation ran 100 thousand time steps. We calculated the force by multiplying the shear pressure output by the area of the surfaces. Figure 10 shows the results of this calculation.

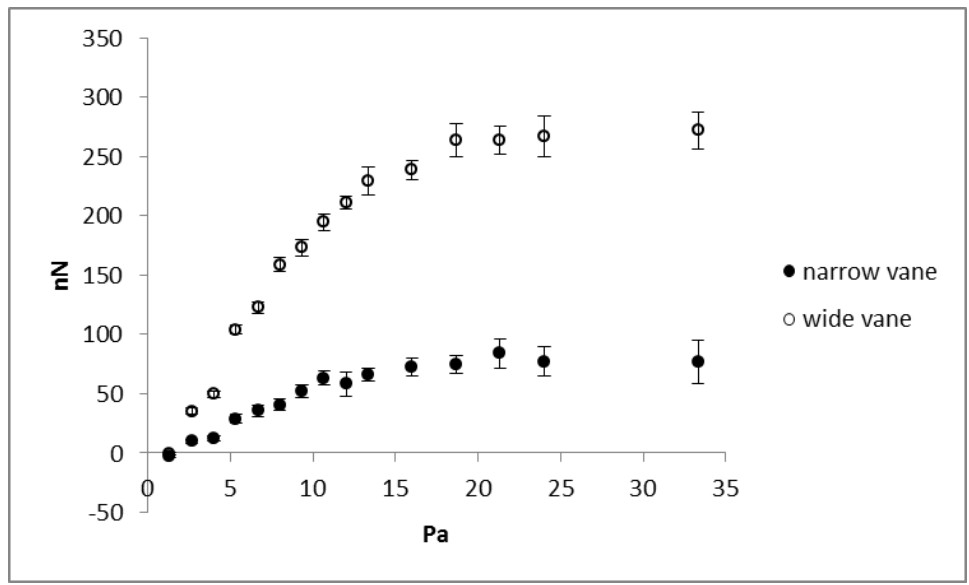

Fig 10. Thermal Creep Force (simulation). The narrow vane is 8mm by 16mm and the wide vane is 16mm by 16mm. The temperature difference on both vanes is 9K. The characteristic length of the temperature gradient on the surface is 3.5mm. The accommodation coefficient is 0.6. The range bars are the length of the standard deviation of the 10 segments of the trial at the reported pressure. Error bars that are not visible are hidden behind the identifying symbol. A positive force is directed from the hot side to the cold side. We believe the results at 1.33 Pa to be spurious.

It is interesting to note that the values are negative at 1.33 Pa (negative meaning pointing towards the hot side so that the radiometer would rotate with the black side leading). For the wide vane, the magnitude of the force is less than the standard deviation.

For the narrow vane though, the force is -2.57 nN with a 1.63 nN standard deviation. None of the 10 time segments of the trial reported a positive force. This result is contrary to the experiment and theory, but similar results have been found by other researchers using the DSMC method to find thermal creep forces. [14] [15] To test the validity of this result we reran the simulation at 1.33 Pa with the narrow vane in the larger simulation space used previously for the wide vane. The reported force was 2.31 nN in the positive direction (pointing toward the cold side). This demonstrates the result is sensitive to the size of the simulation space at 1.33 Pa where there was a single mean free path between the vane and the boundary of the simulation space and the negative results in our trials are not to be trusted. We reran the simulation in the same manner at 6.67 Pa where there were five mean free paths between the vane and the boundary of the simulation space. The reported force was 36.61 nN in the positive direction which compares favorably with the previously determined force of 35.55 nN in the positive direction with a 9.64 nN standard deviation. This suggests the results are stable with respect to the size of the simulation space when there are multiple mean free paths between the vane and the boundary of the simulation space. The trials reported here at 2.67 Pa and above include more than one mean free path between the vane and the boundary of the simulation space. We also ran trials at selected pressures with a higher accommodation coefficient. A higher accommodation coefficient resulted in larger force values as expected.

**B.     DRAG FORCE SIMULATIONS**

We ran drag force simulations identical to the thermal creep simulations with two exceptions. Firstly, we modeled the vanes as a single rectangular object with a temperature of 293 K. Secondly, collisions between the vane and particles were calculated as if the vane were rotating at 10 radians per second around an axis 4 mm from the edge of the vane though the vane was not actually rotating in the simulation space. Figure 11 shows the results of these trials.

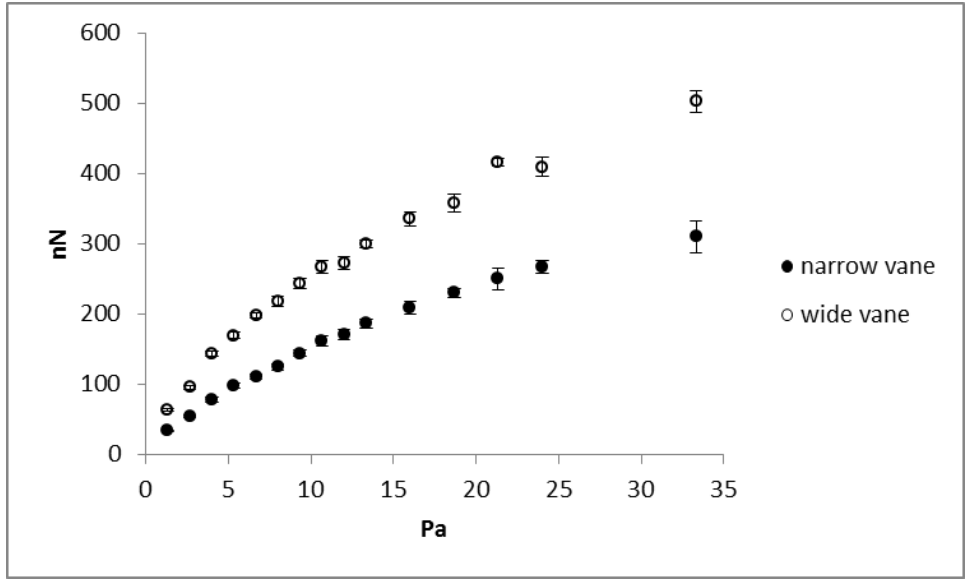

Fig 11. Drag Force at 10 rad/s (simulation). The narrow vane is 8mm by 16mm and the wide vane is 16mm by 16mm. Particle-surface collisions are calculated as if the vane is rotating about a point 4mm beyond its edge. The accommodation coefficient is 0.6. The range bars are the length of the standard deviation of the 10 segments of the trial at the reported pressure. Range bars that are not visible are hidden behind the identifying symbol.

## C. ANGULAR SPEED

We calculated the angular speed predicted by the simulations in the same manner as we calculated the angular speed predicted by theory,

$$\omega = \omega_0 \frac{r_{tc} F_{TC}}{r_d F_D}, \tag{17}$$

where $\omega_0$ is the angular speed in the simulation. This method is possible because the drag force is linearly proportional to velocity at these pressures as seen in equation (14) and both torques scale linearly with the number of vanes on the radiometer. Figure 12 reports the angular speed predicted by the simulations. We omitted the results at 1.33 Pa because of the questionable nature of the results of our thermal creep simulations at that pressure.

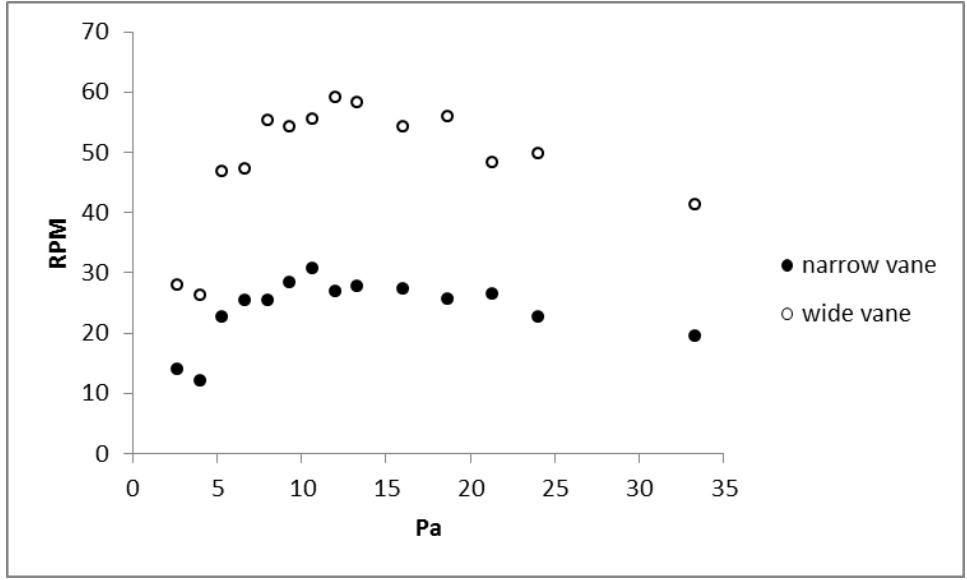

Fig 12. Angular Speed (simulation). The narrow vane is 8mm by 16mm and the wide vane is 16mm by 16mm. Both vanes have a 4mm stem connecting them to the spindle axis. The temperature difference on both vanes is 9K. The characteristic length of the temperature gradient on the surface is 3.5mm. The accommodation coefficient is 0.6. Positive angular speed represents rotation with the cold side leading.

# V.    COMPARISION OF EXPERIMENT, THEORY, AND SIMULATIONS

In both the simulations and theory of the thermal creep force the force rises as a function of pressure at very low pressures then becomes almost constant as a function of pressure at relatively higher pressures. Quantitatively the differences at most pressures are less than an order of magnitude. A qualitative difference does stand out though. The simulations show the force on the narrow vane and the wide vane asymptotically approaching different values as the pressure increases while the theory has the force on the vanes asymptotically approaching similar values. Figure 13 compares the simulation results and theory predictions at the range of pressures examined. We omitted the simulation data points at 1.33 Pa because of the questionable validity of our simulation parameters at that pressure.

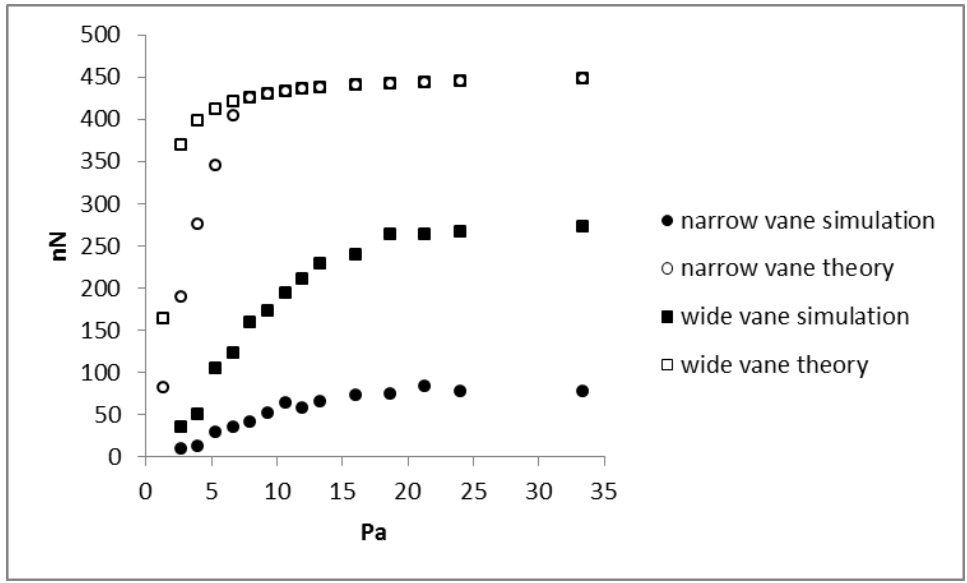

Fig 13. Thermal Creep Force (simulation and theory). The narrow vanes are 8mm by 16mm and the wide vanes are 16mm by 16mm. The temperature difference on the vanes is 9K. The characteristic length of the temperature gradient on the surface is 3.5mm. The accommodation coefficient is 0.6. Positive force is directed from the hot side to the cold side.

In both the simulations and the theory of the drag force the force raises with pressure but the rate of increase decreases with pressure. The change in the rate of increase is more pronounced in the theory than the simulations. Quantitatively the agreement is very good around 13.33 Pa and the differences are less than a factor of two at all pressures examined. Figure 14 compares the simulation results and theory predictions at the range of pressures examined.

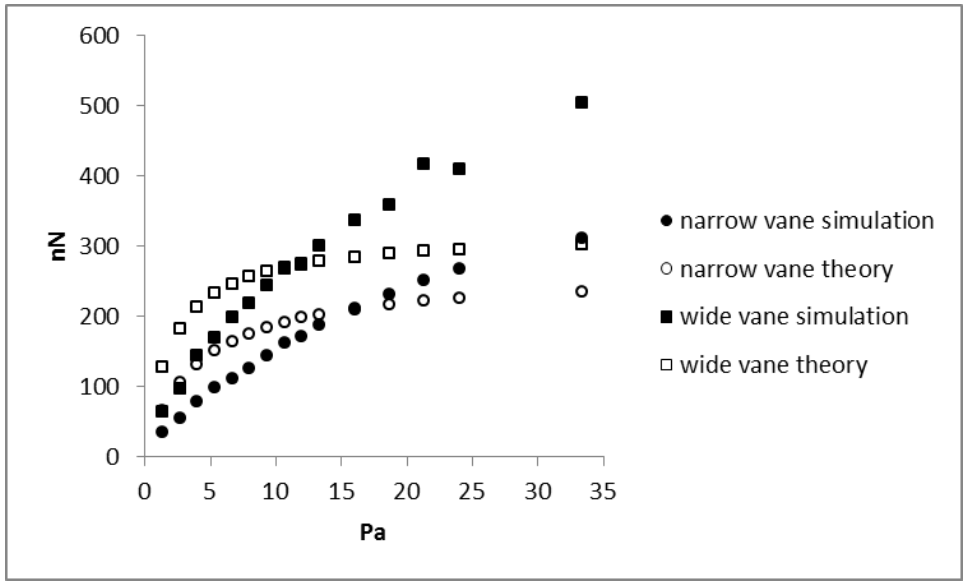

Fig. 14. Drag Force (simulation and theory). The narrow vanes are 8mm by 16mm and the wide vanes are 16mm by 16mm. The vanes have a 4mm stem connecting them to the spindle axis. They are rotating at 10 rad/s. The accommodation coefficient is 0.6.

The experimental results, simulation predictions, and theory predictions of the angular speed of both radiometers all have a peak as a function of pressure between 2.67 Pa and 12 Pa. Quantitatively the differences at most pressures are less than an order of magnitude. The theory is better than the simulations at predicting the experiment results for the narrow vane and the opposite is true for the wide vane. The decrease in the angular speed at relatively higher pressures is more pronounced in the experiment than in the simulation or theory predictions. The difference between the asymptotic force values on the narrow vane and the wide vane in the thermal creep simulations causes the simulation prediction of the angular speed to be greater for the wide vane than the narrow vane. The theory prediction has the narrow vane faster than the wide vane at most pressures. The experimental results qualitatively match the theory predictions in this regard. Figures 15 and 16 compare the experimental results and simulation and theory predictions at the range of pressures examined.

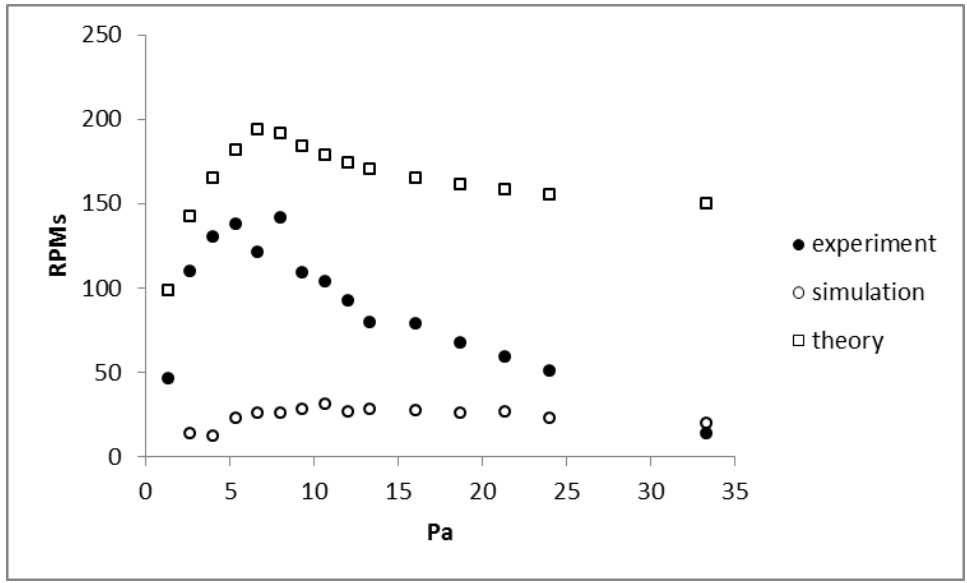

Fig. 15. Narrow Vane Angular Speed (experiment, simulation, and theory). The vanes are 8mm by 16mm. The radiometer has 4 vanes attached to the spindle by 4mm stems. The temperature difference on the vanes is 9K. The characteristic length of the temperature gradient on the vanes is 3.5mm. Positive angular speed represents rotation with the cold side leading.

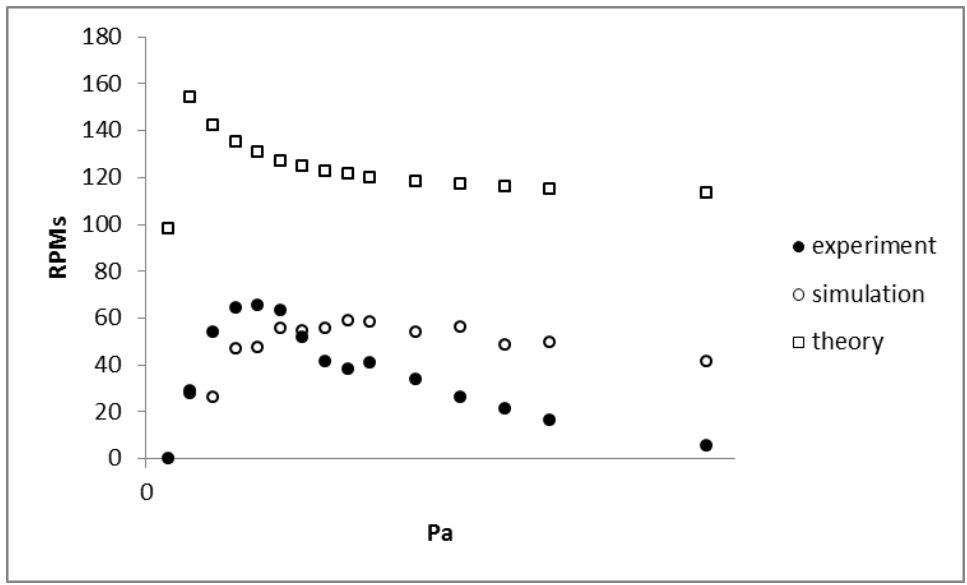

Fig 16. Wide Vane Angular Speed (experiment, simulation, and theory). The vanes are 16mm by 16mm. The radiometer has 4 vanes attached to the spindle by 4mm stems. The temperature difference on the vanes is 9K. The characteristic length of the temperature gradient on the vanes is 3.5mm. Positive angular speed represents rotation with the cold side leading.

# VI.    CONCLUSION

We have investigated the thermal creep shear force through experiment, kinetic theory calculations, and the DSMC simulations. All three investigations show the force acting from the hot to the cold side of the surface. The theory and the simulations both show the force rising with pressure and asymptotically approaching a constant value though the theory shows the force on both vanes approaching the same value while the simulations show the force on the two vanes approaching different values. The theory and the simulations both show the decrease in the radiometer's angular speed as a function of pressure at relatively higher pressures to be the result of increasing drag force rather than decreasing thermal creep force.

Left unexplained by our investigation is why the theory is better at predicting the experimental results for the narrow vane and the simulation results are better at predicting the experimental results for the wide vane. This suggests that at least two of the three methods of investigation, while good, can be further improved upon. There are potential areas of future work which would provide additional insight in all three methods of investigation. In our experiment, we measured RPMs which are the result of all forces on the moving radiometer. Additional insight on the thermal creep force in particular could be achieved by static force measurements on a fixed vane. In our theory, we made a quasi-static assumption and made assumptions about the temperature distribution of the gas. Adding dynamic effects to the theory and experimentally or in simulation directly observing the temperature distribution in the gas and incorporating the results could improve our theory. In our simulations, the thermal creep force and drag force were determined in separate trials. Trials with both a rotating vane and a temperature gradient could reveal any second order effects from the interaction of drag and thermal creep.

# ACKNOWLEDGMENTS

The authors wish to thank Jeffery Catterlin, Steven Jacobs, Aaron Lopez, Jose Lopez, and Rebecca Marvin for running tests on various radiometers. We would also like to acknowledge valuable comments made by the anonymous reviewers. D.W. wishes to thank Michael Gallis and Steve Plimpton for technical assistance with SPARTA and the Naval Postgraduate School Research Computing Group and Department of Defense High Performance Computing Modernization Program for the use of their equipment.